\documentclass[a4paper,11pt]{article}
\pdfoutput=1 

\usepackage{jinstpub} 

\title{Performance study of SKIROC2/A ASIC for ILD Si-W ECAL}

\author[a,1]{T. Suehara,\note{Corresponding author.}}
\author[a]{I. Sekiya,}
\author[b]{S. Callier,}
\author[c]{V. Balagura,}
\author[c]{V. Boudry,}
\author[c]{J-C. Brient,}
\author[b]{C. de la Taille,}
\author[a]{K. Kawagoe,}
\author[d]{A. Irles,}
\author[c]{F. Magniette,}
\author[c]{J. Nanni,}
\author[d]{R. P\"oschl,}
\author[a]{T. Yoshioka,}
\author{and ILD SiW-ECAL group}

\affiliation[a]{Kyushu University,\\744 Motooka, Nishi-ku, Fukuoka, 819-0395 Japan}
\affiliation[b]{Omega Microelectronics Center, \\ \'Ecole polytechnique, 91128 Palaiseau C\'edex, France}
\affiliation[c]{Laboratoire Leprince-Ringuet, \\ \'Ecole polytechnique, 91128 Palaiseau C\'edex, France}
\affiliation[d]{Laboratoire de l'Acc\'el\'erateur Lin\'eaire, \\Centre Scientifique d'Orsay, 91898 Orsay C\'edex, France}

\emailAdd{suehara@phys.kyushu-u.ac.jp}

\abstract{
The ILD Si-W ECAL is a sampling calorimeter with tungsten absorber and highly segmented silicon layers for the International Large Detector (ILD), one of the two detector concepts for the International Linear Collider. SKIROC2 is an ASIC for the ILD Si-W ECAL. To investigate the issues found in prototype detectors, we prepared dedicated ASIC evaluation boards with either BGA sockets or directly soldered SKIROC2. We report a performance study with the evaluation boards, including signal-to-noise ratio and TDC performance with comparing SKIROC2 and an updated version, SKIROC2A.
}

\proceeding{Calorimetry for the High Energy Frontier (CHEF2017)
  Oct. 2017\\
  Lyon}

\begin{document}
\maketitle
\flushbottom

\section{Introduction}

The International Linear Collider (ILC) is a future electron-positron collider with
center-of-mass energy of 250 - 1000 GeV, for precise measurements of Higgs and
electroweak properties and searches for new particles.
The International Large Detector (ILD) is one of two detector concepts being developed for ILC.
The key feature of ILD is `particle flow', which is a method to measure
jet energy precisely, by separating each particle on a jet. This requires
high granularity in the calorimeter, especially on the electromagnetic calorimeter (ECAL).
Silicon-tungsten ECAL (SiW-ECAL) is a suitable solution for the high-granular calorimetry.
It is a sandwich calorimeter with tungsten absorber and silicon pad detectors
with 5x5 mm cells. ASICs should be embedded between the layers
to realize the readout of $\sim10^8$ channels in total.

\begin{figure}[htbp]
\centering
\includegraphics[width=.4\textwidth]{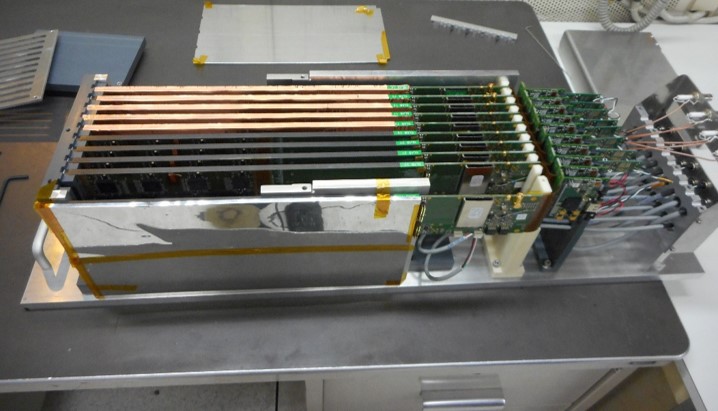}
\caption{\label{fig:proto} A SiW-ECAL prototype used for a test beam.}
\end{figure}

A SiW-ECAL technological prototype\cite{ecal1}\cite{ecal2} is being developed to test and demonstrate
the technology to be used for ILD ECAL. Fig.~\ref{fig:proto} shows the 
setup of technological prototype for a test beam.
Each silicon layer is equipped with four silicon pad sensors and 16 ASICs
for the readout of 1024 channels. The test beam campaigns have been largely successful
and we could demonstrate the readiness of SiW-ECAL for ILD, but 
we found several issues related to the ASICs and PCBs such as fake triggering
due to pedestal shifts and misconfiguration of the threshold tuning function.
To investigate these issues, we investigate the property of the ASICs more precisely,
using a dedicated setup.

\section{SKIROC2 and testboard}

\begin{figure}[htbp]
\centering
\includegraphics[width=.6\textwidth]{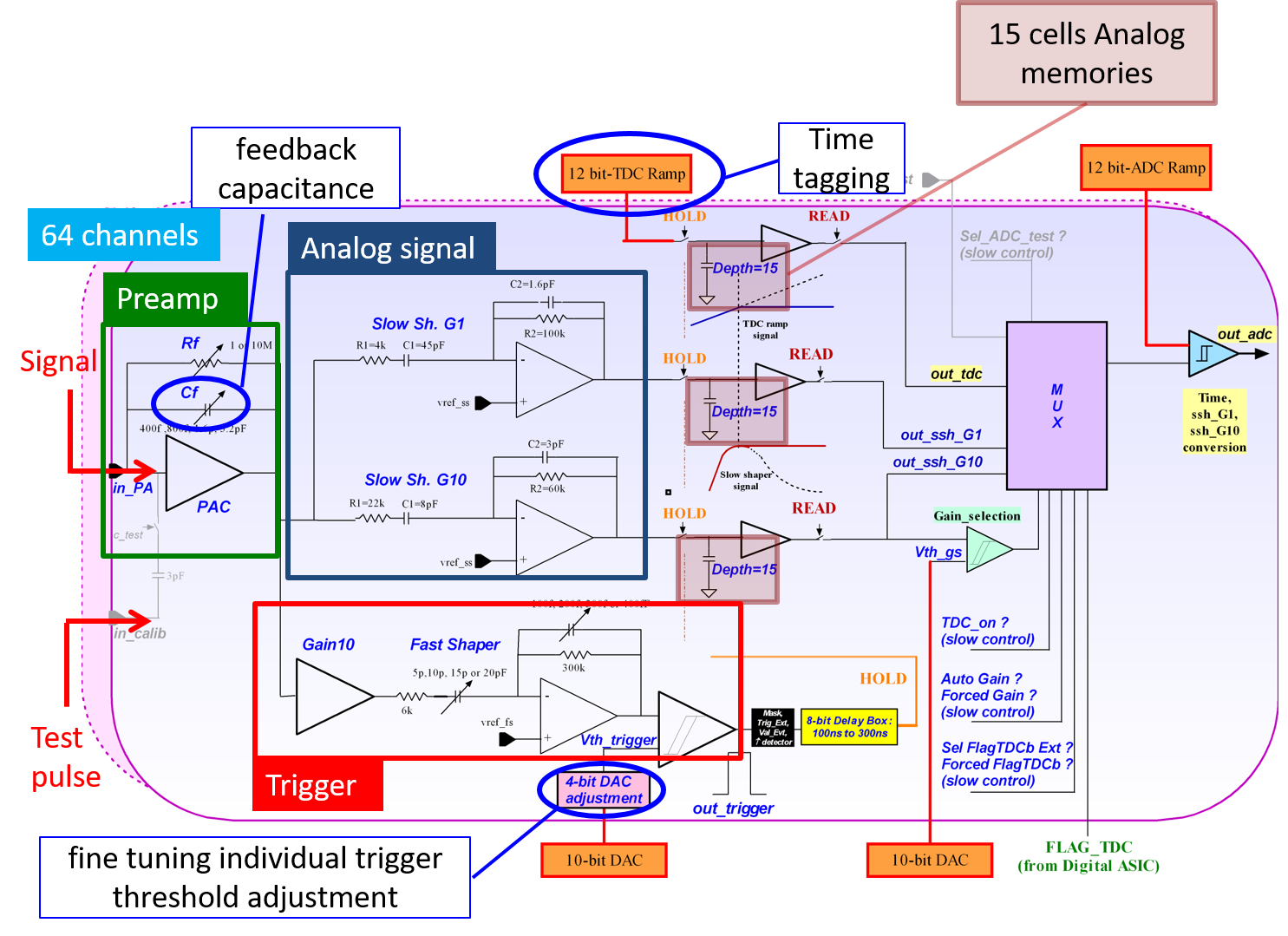}
\caption{\label{fig:schematic} The schematic of the analog part of SKIROC2.}
\end{figure}

SKIROC2 (Silicon Kalorimeter Integrated Read-Out Chip 2) \cite{skiroc2} is an ASIC developed
for the SiW-ECAL readout by Omega group in France. Fig~\ref{fig:schematic} shows
the overview of the analog part of SKIROC2. It has 64 input channels, amplified by
a preamplifier with variable gain and three shaper amplifiers. Two are slow shapers
for ADC measurements with high and low gain, and the other is a fast shaper for
trigger generation. The trigger threshold can be controlled both
globally and locally channel by channel, but the local threshold control
has a too small dynamic range to use in SKIROC2 due to a mistake in the development.
The trigger holds the voltage of the two slow shapers with a tunable delay, and also holds
the voltage swept with the bunch clock trigger to obtain timing information.
There is a 15-deep analog memory to store the voltages during the ILC bunch train
(1312 bunches in about 1 milliseconds in the baseline design).
After the bunch train, the acquisition is stopped and the ASIC changes its state
to readout mode. It uses an ADC with a multiplexer to digitize charge and timing information
of 64 channels and sends the digital data by a serial communication channel.

The most distinct feature of the SKIROC2 chip is the power pulsing functionality, which
partially switches off power when not needed, e.~g.~the amplifiers during the readout period,
significantly reducing power consumption and therefore heat dissipation.
However, this can cause instability of the performance so we should check it carefully
with power pulsing control.

Due to several problems, the Omega group modified the design of SKIROC2 to produce
a bug-fixed version of SKIROC2A. Several modifications are implemented such as
a fix to the dynamic range of local threshold control, change of the TDC voltage sweep shape,
some improvements to the treatment of ground, etc. We compared the performance
of SKIROC2 and SKIROC2A with a setup described below.

\begin{figure}[htbp]
\centering
\includegraphics[width=.4\textwidth]{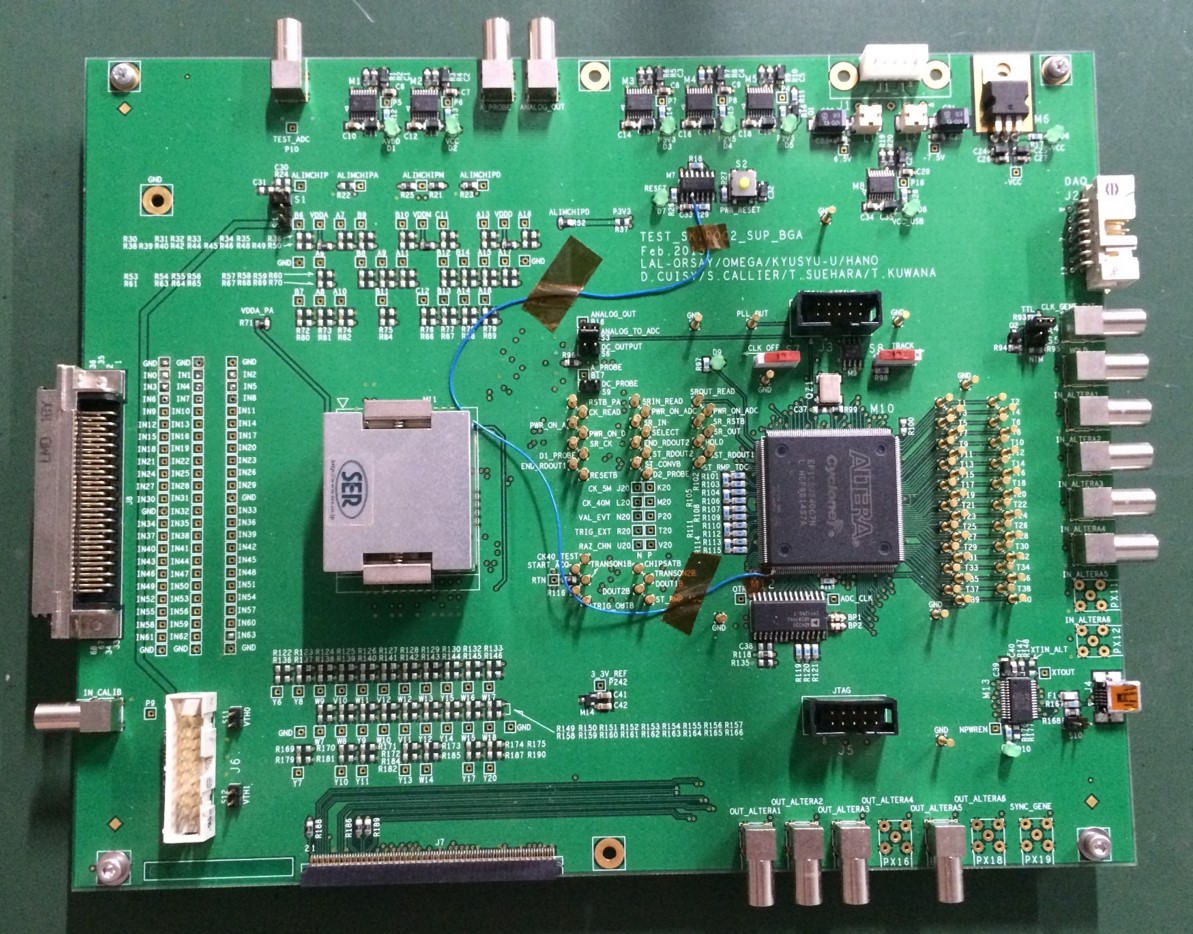}
\caption{\label{fig:testboard} The SKIROC2 testboard with a BGA socket.}
\end{figure}

Figure \ref{fig:testboard} shows a picture of a board for evaluation of SKIROC chips.
The board was developed based on the board with mostly the same function, but for
QFP-packaged SKIROCs. Due to the constraints of size and thickness, we changed the
configuration to use BGA packaged ASICs, so the board was modified by the Kyushu University group to accommodate
BGA SKIROC2. Both socket and soldered versions are available. 

For the operation of SKIROC2, Omega provided firmware and LabView software.
It can control slow control of SKIROC2 as well as getting data from it.
We developed a standalone C++ data acquisition software which can save data
with compatible format to the DAQ of the ECAL prototype.
For power pulsing, the testboard can control one of four power pulsing
channels (analog, digital, DAC and ADC) by an external signal.
In this measurement we use only analog power pulsing and all other
channels are switched on at all times.
Due to the external capacitors on the testboard connected to some ASIC pins,
the recovery time after switching on the analog power is slower than that
in the prototype, more than several milliseconds. Currently we switch on
the power 10 msec prior to the test signal.

\section{Measurements}

Here we show results of several measurements done on SKIROC2/A and the testboard.
All results are obtained with the feedback capacitance of preamplifier set to 1.2 pF,
which gives good linearity up to $\sim 250$ MIPs. The feedback capacitance determines
the gain, and 6.0 pF (with 5 times less gain) is nominal for the real ILC, but
we applied 1.2 pF since we use it for the test beam, so we can directly compare
our results with those from the test beam. With 6.0 pF the signal-to-noise ratio is usually worse.

\subsection{Control of trigger threshold}

The SKIROC2 has a function to tune the trigger threshold for individual
input channels in addition to a global threshold. However, this has an issue
that the dynamic range of 4-bit DAC for the control is too small that
we can hardly control the threshold of individual channels.
Since this has been fixed in the SKIROC2A, we measured the dynamic range
of the individual threshold control.

We used the charge injection of a 1 MIP equivalent charge (4.2 fC) to each channel
and measuring the trigger efficiency at different settings of the global threshold (S-curve measurement).
We measure the trigger efficiency when setting the 4-bit DAC to 0, 4, 8, 12 and 15
to compare the global DAC number with 50\% efficiency.

\begin{figure}[htbp]
\centering
\includegraphics[width=.35\textwidth]{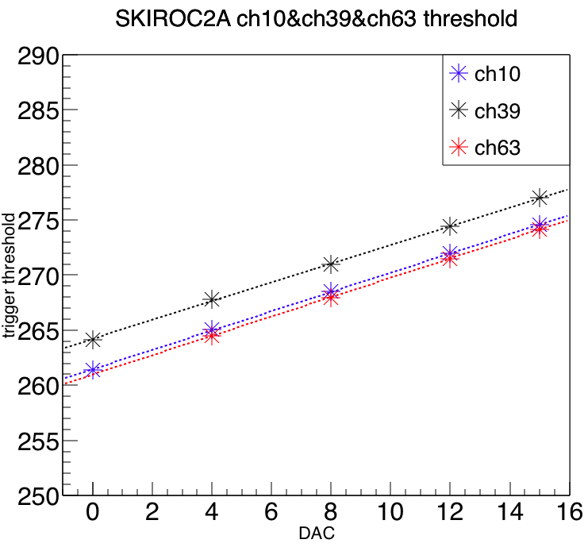}
\caption{\label{fig:threshold} The dynamic range of local threshold control with respect to the global threshold.}
\end{figure}

Figure \ref{fig:threshold} shows the dependence of three sample channels on SKIROC2A.
It shows that the dynamic range of 15 individual DAC units equal to 12.75 global DAC units,
which is 0.13 MIP equivalent, with a very linear response.
This is more than 10 times a larger dynamic range compared to SKIROC2, and it enables us
to use some noisy channels with maximum 0.13 MIP higher trigger threshold.

\subsection{S/N ratio of trigger}

The signal-to-noise (S/N) ratio of the trigger can be obtained via the S-curve technique.
We define the DAC count per MIP as the difference of the DAC values which give 50\% trigger efficiency
for injections of 1 and 2 MIPs,
and the width of the error function fitted to the
threshold scan divided by the DAC value at 50\% as the S/N ratio.

\begin{figure}[htbp]
\centering
\includegraphics[width=.35\textwidth]{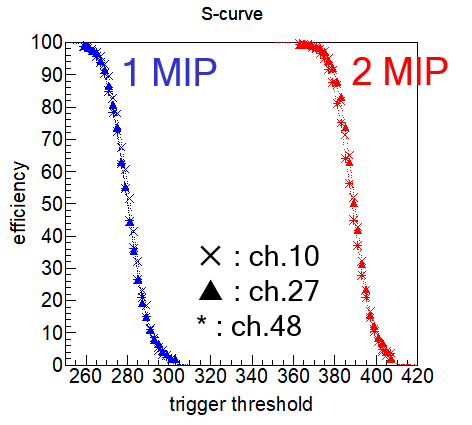}
\qquad
\includegraphics[width=.35\textwidth]{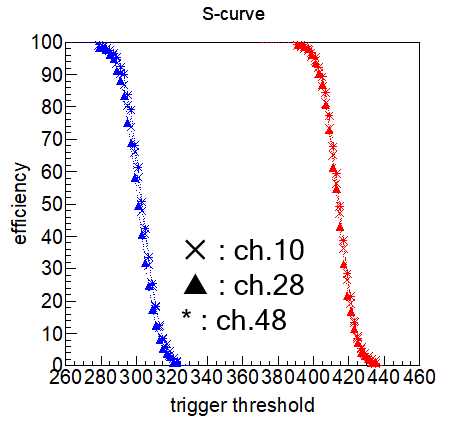}
\caption{\label{fig:trigger} S-curves of trigger threshold with SKIROC2 (left) and SKIROC2A (right).}
\end{figure}

Figure \ref{fig:trigger} shows the S-curve of three channels of SKIROC2 and SKIROC2A, respectively.
This shows the S/N ratio of 12.8 for both of the chips.
This is done with the board with a socket, so it should be better with the soldered board.

\subsection{S/N ratio of ADC}

The S/N ratio of the slow shaper with high gain is checked using the ADC distribution.
We estimated the S/N ratio as the pedestal width fitted by a Gaussian function divided by the gain
calculated as the difference between the response with 1 and 2 MIPs injection.

\begin{figure}[htbp]
\centering
\includegraphics[width=0.9\textwidth]{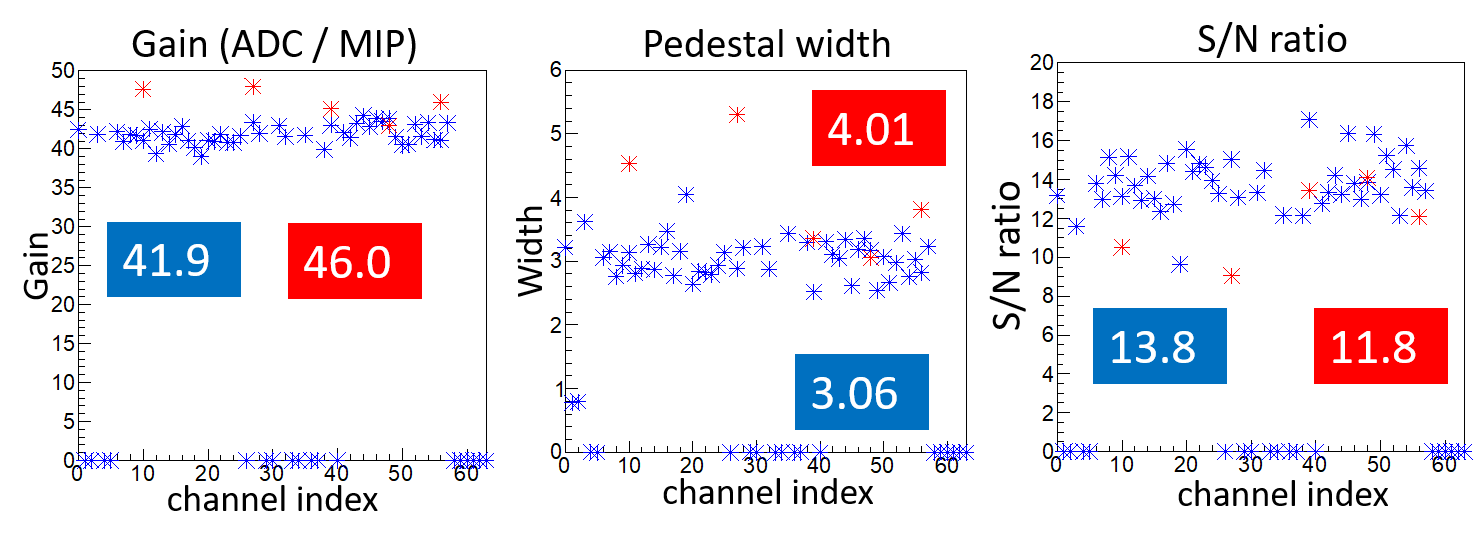}
\caption{\label{fig:adc} Gain, Noise, and S/N ratio of ADC measurements on SKIROC2 and SKIROC2A
with the socket. Blue (red) points and numbers show the results and their average with SKIROC2 (SKIROC2A).}
\end{figure}

Figure \ref{fig:adc} shows the channel variation of gain, width and S/N ratio
with both SKIROC2 and SKIROC2A with the socket. For SKIROC2, we scanned all 64 channels
but it shows several channels could not give output because of problems with some socket connections.
We do not see similar problems on the soldered board.
For SKIROC2A we picked several channels.
Numbers in the plot shows the avarage, which shows a slight difference on S/N ratio, but
we will investigate whether it is a true difference of performance or just a statistical
fluctuation or some issues related to connection or measurements.

The obtained S/N ratio is significantly worse than what we got previously with the soldered
SKIROC2A (which is around 26), but the setup such as the injection pulse shape was
different, so we are still investigating the difference.

\subsection{TDC measurement}

The TDC feature of SKIROC2A is tested for the first time.
We used 5 MIP injection and provided the injection pulse
synchronized to the bunch clock with a variable delay.
By scanning the delay we can get the response of the TDC output
with respect to the delay.

\begin{figure}[htbp]
\centering
\includegraphics[width=.7\textwidth]{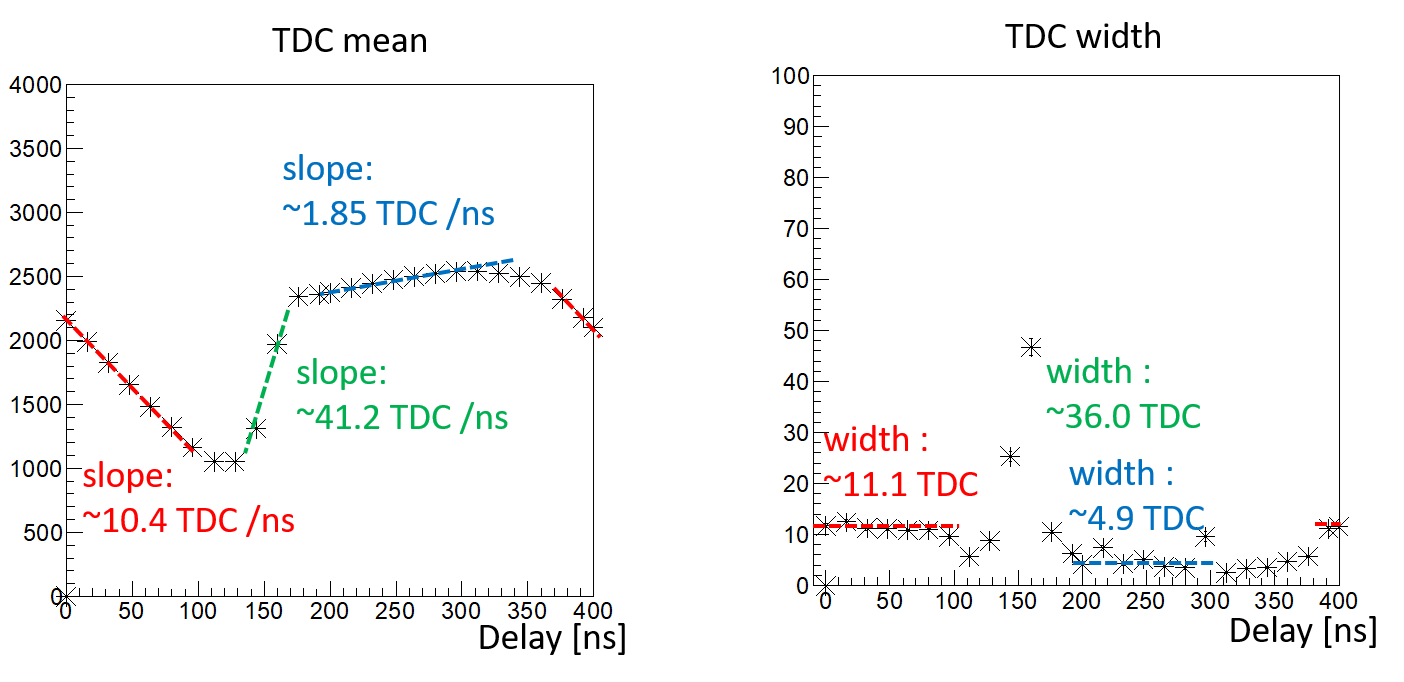}
\caption{\label{fig:tdc} The mean (left) and the width (right) of TDC distribution with SKIROC2A on the soldered testboard.}
\end{figure}

Figure \ref{fig:tdc} shows the central value and width of the TDC
output when scanning the delay with the soldered testboard of SKIROC2A.
It shows the voltage rising and falling in the period of two bunch clocks (400 nsec).
The shape is not triangular as designed in the green and blue regions which should be due to
misconfiguration of the ASIC, but the timing resolution can be estimated from the
red area, which is around 1.1 nsec.

\section{Summary and prospects}

We measured characteristics of SKIROC2 and SKIROC2A using testboards
with a BGA socket and soldered chips with charge injection.
The dynamic range of local threshold control is 0.13 MIPs in SKIROC2A,
which can be used to recover channels with higher noise.
The signal-to-noise ratios of both trigger and ADC have no significant
difference on the measurements with socket.
For the TDC, 1.3 nsec timing resolution is obtained with SKIROC2A.

We plan to measure and check the performance of SKIROC2As with the socket board
before soldering them to the ECAL prototype.
For the next production we need to measure around 90 SKIROC2As, so automatic
measurement will be necessary. We are preparing software to automatically perform S-curve scans and
ADC noise measurements. After finshing the preparation, we will measure
SKIROC2 and SKIROC2A more systematically to compare their performance in more detail.
The system can be expanded for the quality
control of the ILD ECAL production, for which we will need O($10^6$) chips.

\acknowledgments

This work is supported by VLSI Design and Education Center (VDEC), the University of Tokyo
in collaboration with Cadence Design Systems, Inc.
This work is supported by JSPS KAKENHI Grant Numbers JP23000002 and JP17H05407.

\end{document}